\documentclass[aps,prl,twocolumn]{revtex4}
\usepackage{epsfig}
\usepackage{graphicx}
\usepackage{bm}
\begin{document}

\title{Heralded amplification for precision measurements with spin ensembles}

\author{Nicolas Brunner$^1$, Eugene S. Polzik$^2$, and Christoph Simon$^3$}

\affiliation{$^1$ H.H. Wills Physics Laboratory, University of Bristol, Tyndall Avenue, Bristol BS8 1TL, United Kingdom \\ $^2$ Niels Bohr Institute, Danish Quantum Optics Center QUANTOP, Copenhagen University, Blegdamsvej 17, DK-2100 Copenhagen {\O}, Denmark\\ $^3$ Institute for Quantum Information Science and
Department of Physics and Astronomy, University of Calgary,
Calgary T2N 1N4, Alberta, Canada}

\begin{abstract}
We propose a simple heralded amplification scheme for small rotations of the collective spin of an ensemble of particles. Our protocol makes use of two basic primitives for quantum memories, namely partial mapping of light onto an ensemble, and conversion of a collective spin excitation into light. The proposed scheme should be realizable with current technology, with potential applications to atomic clocks and magnetometry.
\end{abstract}

\date{\today}

\maketitle

The ability to measure very small rotations of the collective spins of atomic ensembles lies at the heart of precision measurement applications such as atomic clocks \cite{clocks} and magnetometry \cite{magnet}. In practice the rotations have to have a certain minimum size in order to be detectable, due to noise that can be either intrinsic, e.g. due to the finite number of atoms in the ensemble, or external, e.g. detector dark counts.

There has recently been a lot of progress in the development of probabilistic amplification techniques, particularly in optics. Weak-value based amplification \cite{weak} was used to perform very precise measurements of the optical spin Hall effect \cite{kwiat} and of tiny mirror displacements \cite{howell}. Separate, but similar, developments are the heralded noiseless amplification scheme of optical coherent states \cite{ralph,grangier} and the probabilistic concentration of phase information \cite{marek}. In all of these cases a small effect can be amplified with some probability, where the probability of success decreases with increasing amplification factor.

For weak-value based amplification and heralded noiseless amplification (of weak signals) \cite{ralph,grangier} the amplification factor is approximately $\frac{1}{\sqrt{p}}$, where $p$ is the success probability, which is optimal taking into account the linearity of quantum mechanics \cite{linear}. This implies that these techniques do not help in situations where the measurement error scales like $\frac{1}{\sqrt{R}}$, where $R$ is the number of repetitions of the experiment, because then the gain in amplitude is exactly balanced by the loss in statistics \cite{kwiat}. However, in many cases the measurement error actually scales less favorably with $R$ \cite{kwiat,howell,howellpra,brunner,simon,feizpour}. This is certainly the case for potentially present systematic errors. Another example is if the total measurement time is limited, but the noise has a finite correlation time. In this case by increasing the repetition rate of the experiment one will eventually enter the regime where the noise in subsequent experiments is correlated and thus no longer averages out \cite{feizpour}. In general these probabilistic amplification techniques will help in all situations where the measurement error decreases more slowly than $\frac{1}{\sqrt{R}}$.

It is hence of interest to explore if the concepts of probabilistic amplification can be applied to precision measurements with atomic ensembles. Demonstrations of weak-value measurements involving atomic ensembles were proposed in Refs. \cite{simon,feizpour}. However, these references aim to amplify small light-matter couplings, rather than small rotations of the atomic states by external forces (e.g. due to a magnetic field). Small rotations are the most relevant effect for precision measurements.

Here we propose a heralded amplification scheme for small spin rotations, or equivalently small coherent displacements in phase space. The correspondence between collective states of an atomic or solid-state spin ensemble and harmonic oscillator states is discussed in Refs. \cite{simon,hammerer}. Consider an ensemble of atoms that have levels $g$ and $s$. The atoms all start out in $g$, so that the initial atomic state is $|g\rangle_1 |g\rangle_2 ... |g\rangle_N$, where $N$ is the total number of atoms. Let there be a small unknown rotation
\begin{equation}
|g\rangle \rightarrow |g\rangle + \epsilon |s\rangle
\end{equation}
 for each atom, where $\epsilon$ can in general be complex. Then the new state of the atomic ensemble to first order in $\epsilon$ is
\begin{equation}
|g\rangle_1 |g\rangle_2 ... |g\rangle_N + \epsilon \left( |s\rangle_1 |g\rangle_2...|g\rangle_N + ... + |g\rangle_1...|g\rangle_{N-1}|s\rangle_N \right).
\label{rotated}
\end{equation}
This can be rewritten as
\begin{equation}
|0\rangle + \sqrt{N} \epsilon |1\rangle,
\end{equation}
where $|0\rangle$ and $|1\rangle$ are the normalized symmetric states with zero and one atoms in $s$ respectively.

One can introduce spin operators $\sigma_x=|g\rangle \langle g|-|s\rangle \langle s|$ etc. for each atom, and associated collective spin operators $J_x=\frac{1}{2} \sum_{k=1}^N \sigma_x^{(k)}$ etc. The collective operators satisfy commutation relations $[J_y,J_z]=i J_x$ etc. The state $|0\rangle=|g\rangle_1 |g\rangle_2...|g\rangle_N$ is the eigenstate of $J_x$ with the maximum eigenvalue $N/2$. As long as there are just a few atoms in $s$, one can make the Holstein-Primakoff approximation and introduce canonical variables $X=J_y/\sqrt{N/2}$ and $P=J_z/\sqrt{N/2}$. Under the mentioned condition they satisfy $[X,P]=i$ to excellent approximation \cite{hammerer}. The states and collective spin observables of the atomic ensemble can thus be mapped onto the states and quadrature observables of a harmonic oscillator. In particular the rotated state of Eq. (\ref{rotated}) corresponds to a coherent state $|\alpha\rangle=|0\rangle + \alpha |1\rangle$ of the harmonic oscillator, where we assume that $\alpha=\sqrt{N} \epsilon \ll 1$, because we are interested in the case of very small rotations. Our goal here is to probabilistically amplify the small amplitude $\alpha$.

We now discuss our heralded amplification scheme, which is based on the quantum-optical catalysis protocol of Ref. \cite{lvovsky}. Consider two bosonic modes. One contains the signal, i.e. the coherent state $|\alpha\rangle$; the other one contains an auxiliary single excitation $|1\rangle$. The initial state is thus
\begin{equation}
|\alpha\rangle |1\rangle=|0\rangle |1\rangle + \alpha |1\rangle |1\rangle.
\end{equation}
The two modes are mixed on a beam splitter with small transmission amplitude $t$, where we will assume right away that $\alpha \ll t \ll 1$ and only keep the dominant terms. The state is now
\begin{equation}
|0\rangle |1\rangle + t |1\rangle |0\rangle + \alpha |1\rangle |1\rangle.
\end{equation}
Detecting an excitation in the first mode projects the second mode onto the state
\begin{equation}
t |0\rangle + \alpha |1\rangle = t \left(|0\rangle + \frac{\alpha}{t}|1\rangle \right). \label{ampstate}
\end{equation}
One can see that we have succeeded in amplifying the amplitude $\alpha$ by a factor $1/t$, with a probability of success that is approximately $t^2$. The successful amplification is heralded by the detection of the excitation in the first mode.
The state of Eq. (\ref{ampstate}) is formally identical to the output of the heralded noiseless amplification of Refs. \cite{ralph,grangier}, see in particular Eq. (1) of Ref. \cite{grangier}. There is also a close link to weak-value measurements, see in particular Eq. (2) of Ref. \cite{simon} and the associated discussion. However, the present approach is particularly well suited for metrology or quantum sensing with an ensemble of material spins.

We take the input state $|\alpha\rangle$ to be the state of an atomic ensemble as discussed above. We choose the second mode to be optical.
It would be possible to let the second mode be an atomic mode as well, but the combination of one atomic mode and one light mode is particularly convenient. In order to implement the protocol described above, we have to fulfill three requirements: (i) prepare the optical mode in a single-photon state; (ii) implement a beam splitter operation (with small $t$) between an atomic mode and a light mode; and (iii) detect a single excitation in the first, i.e. atomic, mode.

Requirement (i) is now fairly routinely met in quantum optics experiments. Ref. \cite{lvovsky} is one example for an experiment where this approach was used successfully with a heralded source based on parametric down-conversion; see also Ref. \cite{polzik} where a single-photon source compatible with atomic ensembles is reported. Operations (ii) and (iii) have been demonstrated in experiments on quantum memories \cite{hammerer,lvovskyNP,simonEPJD,kimble,sangouard}. Operation (ii) corresponds to the partial mapping of the single photon onto the atomic ensemble. In the language on quantum memories, we here aim for a storage efficiency $t^2 \ll 1$, which is very easy to achieve in many implementations. Operation (iii) corresponds to reading out the atomic excitation that is stored in the memory. This is most frequently done by converting the atomic excitation into a photon and detecting that photon. This has also been achieved in many experiments \cite{hammerer,lvovskyNP,simonEPJD,sangouard}.

\begin{figure}
\epsfig{file=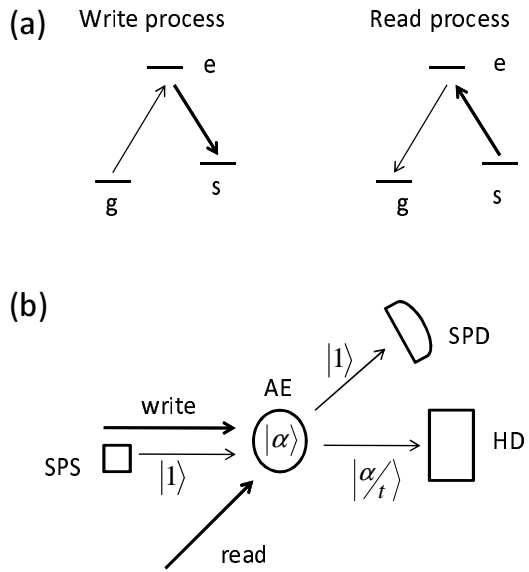,width=0.9\columnwidth}
\caption{Proposed implementation of heralded amplification with an ensemble of three-level systems. (a) Level structure and transitions for the write and read processes. Thin arrows correspond to the write and read photons, whereas thick arrows signify the write and read laser beams. (b) Schematic setup. The atomic ensemble (AE) is originally in the small-amplitude coherent state $|\alpha\rangle$. An auxiliary photon produced by the single-photon source (SPS) is partially stored in the ensemble with the help of the write beam, where $t$ is the amplitude for the storage to occur. The state of the ensemble is subsequently read out with the help of the read beam. If a read photon is detected in the single-photon detector (SPD), then the transmitted light is in the amplified coherent state $|\alpha/t\rangle$, which can be measured by homodyne detection (HD).}
\end{figure}

For example, all of these steps can be implemented using Lambda systems, where there is a third level $e$ that is coupled to both level $g$ and level $s$ by optical transitions, see Figure 1. The initial photon is prepared such that it is in resonance with the $g-e$ transition. It can be stored in the ensemble by applying a ``write'' laser beam on the $g-s$ transition simultaneously with the single photon (and typically propagating in the same direction). The probability amplitude for the absorption to occur can be adjusted for example by varying the number of atoms in the ensemble or the intensity of the write beam. Reading out an atomic excitation works in basically the same way. If a strong ``read'' laser beam is applied on the $g-s$ transition, a collective atomic excitation in $s$ - as described above in the context of Eq. (1) - will be converted into a photon in resonance with the $g-e$ transition. This ``read'' photon will be emitted into a well-defined direction (parallel to the read beam, in the simplest case)  thanks to collective interference of the emission amplitudes from all the atoms in the ensemble \cite{sangouard}, and is thus fairly easy to detect. Note that the read beam can be non-collinear with the incoming photon and the write beam, so that the read photon can be distinguished from the light that is transmitted through the ensemble, see Figure 1.

This last point is important because, if all these steps are completed successfully, the amplified coherent state $|\frac{\alpha}{t}\rangle=|0\rangle+\frac{\alpha}{t}|1\rangle$ is prepared in the transmitted optical mode. In fact, it is possible to detect the state of the light immediately after the partial absorption process. One can then simply post-select the relevant cases, i.e. the cases in which a read photon (and thus an atomic excitation) is detected. The most obvious measurement technique to use in order to see the amplified amplitude is homodyne detection, which is also a well-established experimental technique \cite{lvovsky,hammerer}. 

Let us comment on a few more points of implementation. The single photon source does not have to be perfect. In particular, a vacuum component will simply not give any detections in the read mode (or only detections from the original excitation, which we assume to have too low probability to be detected reliably), and will thus not contribute significantly to the results. Similarly, the read efficiency also doesn't have to be perfect. If an actually present atomic excitation is not detected, this will also simply reduce the success probability of the experiment. One important experimental challenge will be the low success probability necessary for large amplification factors. A single run of a storage and readout experiment with cold atoms typically takes of order 100 ms. If one aims for an amplification factor of 30, corresponding to a success probability of order $10^{-3}$, a single point would thus take about two minutes. This time can be reduced by using room-temperature atoms. The low success rates also require very low dark count rates for the single-photon detector.

In conclusion, we have presented a heralded amplification scheme for small rotations (which we mapped onto small coherent displacements) in an atomic ensemble. The scheme makes use of standard techniques from the fields of quantum optics and quantum memories. Its experimental implementation should thus be possible with current technology.  We believe that the proposed technique has the potential of significantly enhancing precision measurements with atomic ensembles.

This work was supported by an AITF New Faculty Award, an NSERC Discovery Grant, the UK EPSRC, and the European projects Q-ESSENCE and MALICIA.

\end{document}